\documentclass[12pt]{article}

\usepackage{latexsym}

\makeatletter
\@addtoreset{equation}{section}
\makeatother

\newcommand{\eref}[1]{(\ref{#1})}

\def\beq{\begin{equation}}
\def\eeq{\end{equation}}
\def\bea{\begin{eqnarray}}
\def\eea{\end{eqnarray}}
\def\ra{\rightarrow}
\def\a{\alpha}
\def\b{\beta}
\def\g{\gamma}
\def\l{\lambda}
\def\f{\phi}

\def\e{\varepsilon}

\def\G{\Gamma}

\def\D{\partial}

\def\um{\underline{m}}
\def\un{\underline{n}}
\def\ua{\underline{\a}}

\begin{document}

\begin{titlepage}

\begin{flushright}
Preprint DFPD 97/TH/44\\
October 1997\\
\end{flushright}

\vspace{2truecm}

\begin{center}

{\bf \Large Covariant actions for $N=1,D=6$  Supergravity theories 
with chiral bosons}

\vspace{1cm}

{Gianguido Dall'Agata}\footnote{dallagata@pd.infn.it}, {Kurt 
Lechner}\footnote{kurt.lechner@pd.infn.it} and Mario 
Tonin\footnote{mario.tonin@pd.infn.it}

\vspace{1cm}

{\it Dipartimento di Fisica, Universit\`a degli Studi di Padova,

\smallskip

and

\smallskip

Istituto Nazionale di Fisica Nucleare, Sezione di Padova, 

Via F. Marzolo, 8, 35131 Padova, Italia}

\vspace{1cm}

\begin{abstract}

We show that the recently found covariant formulation for chiral 
$p$--forms in $2(p+1)$ dimensions with $p$ even, can be naturally
extended to supersymmetric theories. We present the general method
for writing  covariant actions for chiral bosons and 
construct, in particular, in six dimensions covariant actions for one
tensor supermultiplet, for pure supergravity and for supergravity 
coupled to an arbitrary number of tensor supermultiplets.
	
\end{abstract}

\end{center}
bosons

\end{titlepage}

\newpage

\baselineskip 6 mm


\section{Introduction}

Many models in $D$ dimensions, with $D/2$ odd, contain chiral bosons 
i.e. $p$--form gauge fields (scalars in $D = 2$) with self--dual or 
anti--selfdual field strength. The description of the dynamics
of such theories through Lorentz--invariant
actions has been an unsolved problem for long time \cite{MS}. 
Previous attempts to face this problem were based on 
non manifestly covariant actions \cite{nmcov,PS} or on formulations 
involving an infinite set of auxiliary fields \cite{infin}. 
On the other hand Siegel's approach to two dimensional chiral bosons 
\cite{Siegel} admits only rather problematic extensions to 
$D > 2$ \cite{armeni}.

Recently a new manifestly Lorentz--invariant 
approach \cite{PST,PST3,M5} for chiral $p$-forms in $D = 2(p+1)$ with $p$ 
even has been proposed. This approach is based on a single auxiliary
scalar field and allows to write actions which are manifestly
invariant under Lorentz transformations and under two 
new bosonic local symmetries. The gauge fixing of these symmetries 
allows to remove the auxiliary field and to eliminate  half of the 
physical degrees of freedom of the $p$--form so that its 
field strength becomes (ant)--selfdual in the free case, or satisfies 
a generalized selfduality condition in the interacting case.
Moreover, this approach reproduces on one hand, through an appropriate
gauge fixing, the non manifestly covariant approach of 
\cite{PS} and it is, on  the other hand, related to the 
approach in \cite{infin} in that it provides a consistent truncation of its 
infinite set of auxiliary fields.

The new approach has been first introduced in \cite{PSTplb} to 
reformulate the four--dimensional Maxwell theory
without sources in such a way that its invariance under 
electric/magnetic duality and under Lorentz transformations is manifest;
the coupling to electric and magnetic sources was performed in \cite{berk}.
The approach allowed also to rewrite the  heterotic string effective 
action of \cite{M5alt} in a form such that duality symmetry is manifest
\cite{PSTprd}. 
The method appeared also perfectly suitable \cite{PST,PST3} for
providing lagrangian formulations for theories with chiral $p$--forms. 
It has been used, in particular, to obtain a manifestly Lorentz invariant
and $\kappa$--invariant action for the eleven--dimensional
$M$--theory five--brane \cite{M5} and to obtain a new 
formulation for  Green--Schwarz heterotic strings which involves chiral 
bosons instead of heterotic chiral fermions \cite{stringa}.

An equivalent non--manifestly covariant formulation of the 
$M$--theory five--brane has been given in \cite{APPS}. 

This theory has also
been worked out in \cite{HS} on the basis of a purely geometrical 
doubly supersymmetric approach which does, however, not furnish
a Lagrangian formulation but only the equations of motion.
Subsequently in ref. \cite{BNLPST} it has been shown that
this formulation, at the level of the field equations, 
is equivalent to the ones presented in \cite{M5} and \cite{APPS}.

The coupling of all these models with chiral bosons to gravity can be
easily achieved since the approach is manifestly covariant
under Lorentz transformations; as a consequence it is
obvious that the two above mentioned  bosonic symmetries, which are
a crucial ingredient of the new approach, are compatible with
diffeomorphism invariance. Therefore, to establish the general
validity of the approach, it remains to establish its compatibility
with global and local supersymmetry. This is the aim of the
present paper.

Chiral $p$--forms are, in fact, present in many supersymmetric and 
supergravity models in two, six and ten dimensions.
The covariant action for the bosonic 
sector of $D = 10$, $IIB$ supergravity has already been described 
in \cite{IIB}.

The problem we address is if one can deduce the dynamics of
supersymmetric models with chiral bosons from supersymmetry
invariant actions which respect also the new bosonic 
symmetries of the covariant approach. There is already some
evidence that supersymmetry is compatible with the new bosonic 
symmetries in a rather natural way. Indeed, on one hand in refs.
\cite{PSTplb,PST2} a four dimensional model has 
been considered where these bosonic symmetries and rigid 
supersymmetry are both simultaneously present:
this was obtained by a simple modification of the supersymmetry 
transformations of the fermions, which vanishes on--shell.
On the other hand, in the $M$--five--brane action invariance under 
$\kappa$--symmetry was achieved in a very natural way.

The present paper deals, in particular, with the problem of 
constructing covariant 
actions for supersymmetric and supergravity models with chiral 
bosons in {\it six} dimensions. 

In six dimensions there are four kind of multiplets: the supergravity 
multiplet, the tensor multiplet, the vector multiplet and the 
hypermultiplet. The supergravity multiplet and the tensor multiplets 
contain a chiral two--form with anti--selfdual and selfdual field 
strength respectively.
The on--shell supersymmetry transformation rules as well as the field 
equations for these supermultiplets are well known \cite{sugra6d}.
Moreover, in 
\cite{DFR} the group--manifold action for $D = 6$ pure supergravity has been 
obtained. Usually the group manifold action, when restricted to ordinary
space--time, gives rise to an action for the component fields which
is invariant under supersymmetry. In the presence of chiral bosons,
however, the component action obtained in this way fails to be susy
invariant. Proposals for covariant actions for  pure supergravity in $D=
6$ have been made in \cite{armeni}, which constitutes a generalization
of Siegel's approach \cite{Siegel} from two to six dimensions, and 
in \cite{Sok} where a harmonic superspace lagrangian 
approach is used, which involves, however, an infinite number 
of auxiliary fields. The same kind of problems arises also in $D=10$,
$IIB$ supergravity which involves a four-form with selfdual field
strength\footnote{Also for $D = 10$, $IIB$ supergravity
the on shell susy transformation rules and the field equations are well known 
\cite{sugraIIB} and a group manifold action exists \cite{genIIB}; 
an action based on the Siegel approach has been proposed in
ref. \cite{armeni}.}

In the present paper we shall show that the new covariant approach 
for chiral $p$--forms allows to obtain in a natural and elegant way, 
covariant and supersymmetric actions for six dimensional 
models with chiral bosons which 
involve the supergravity multiplet and/or tensor 
multiplets. For completeness in section two we review the 
covariant approach for a chiral two--form in six dimensions.
In section three we illustrate our technique for writing supersymmetric
covariant actions with chiral bosons in the case of a single free tensor 
multiplet in flat space. This simple example appears rather instructive
in that it exhibits all principal features of our technique.
Along the lines of this example we construct in section four the action for 
pure $N = 1$, $D = 6$  supergravity and present in section five  
the action for the more general case of $N = 1$, $D = 6$ supergravity coupled 
to an arbitrary number, $n$, of tensor supermultiplets. 
The couplings of these multiplets with 
hypermultiplets and vector multiplets will be briefly 
discussed in the concluding section six.

The general strategy developed in this paper extends in a rather
straightforward way to two and ten dimensions. Particularly interesting 
is the case of $IIB$, $D = 10$ supergravity whose covariant 
action we hope to present elsewhere.

\section{Chiral bosons in six dimensions: the general method}

In this section we present the method for a chiral boson
in interaction with an external or dynamical gravitational
field in six dimensions. To this order we introduce sechsbein one--forms
$e^a = d x^m {e_m}^a(x)$. With  $m,n =0,\ldots,5$ we indicate 
curved indices and with $a,b=0,\ldots,5$ we indicate tangent
space indices, which are raised and lowered with the flat
metric $\eta_{ab}=(1,-1,\cdots,-1)$.
We define the tangent space components of a generic 
$p$--forms, $\phi_p$, according to  
\beq\label{dec}
\phi_p={1\over p!}e^{a_1}\cdots e^{a_p}\phi_{a_p \cdots a_1},
\eeq
where the wedge product between forms here and in the following 
will always be understood.

To consider a slightly more general self-duality condition for
interacting chiral bosons we introduce the two-form potential 
$B$ and its generalized curvature three--form $H$ as
$$
H=dB+C\equiv {1\over 3!}e^a e^b e^c H_{cba},
\label{forms}
$$
where $C$ is a three-form which depends on the fields to which 
$B$ is coupled, such as the graviton, the gravitino and so on, 
but not on $B$ itself. The free (anti)self--dual boson 
will be recovered for $C=0$ and $e_m{}^a=\delta_m{}^a$.

The Hodge--dual of the three--form $H$ is again a three--form
$H^*$ with components
$$
H^*_{abc} = \frac{1}{3!} \e_{abcdef} H^{def}.  
$$
The self--dual and anti self--dual parts of $H$ are defined respectively
as the three--forms
$$
H^{\pm} \equiv \frac{1}{2} (H \pm H^*).
$$
The equations of motion for interacting chiral bosons in supersymmetric 
and supergravity theories, as we will see in the examples worked out in
the next sections, are in general of the form
\beq
\label{eqm}
H^{\pm}=0,
\eeq
for a suitable three--form $C$ whose explicit expression is usually 
determined by supersymmetry.

To write a covariant action which gives eventually rise to \eref{eqm}
we introduce as new ingredient the scalar auxiliary field $a(x)$
and the one--forms
\bea
u&=&da\equiv e^a u_a\\
v&=&{1\over \sqrt{-u^2}}\, u\equiv e^a v_a.
\eea
In particular we have $v_a={u_a\over \sqrt{-u^2}}$ and $v_av^a=-1$. 
Using the vector $v^a$ to the 
three--forms $H,H^*$ and $H^\pm$ we can then associate two-forms $h,h^*$ 
and $h^\pm$ according to 
$$
h_{ab}=v^cH_{abc}, \qquad h={1\over 2} e^a e^b h_{ba},
$$
and similarly for $h^*$ and $h^\pm$.

The action we search for can now be written equivalently in one of the
following ways
\bea
\label{S0}
S_0^\pm &=& \pm\int \left(v h^{\pm} H + {1\over 2} dB C\right)\nonumber\\
           &=& \pm\int \left(v h^{\pm} H + {1\over 2} H C\right)\nonumber\\
        &=& \int d^6x\sqrt{g}\left({1\over 24}H_{abc}H^{abc} 
         +{1\over 2}h_{ab}^\pm h^{\pm ab}\right) \pm\int {1\over 2}dBC
            \nonumber\\
       &=& {1\over 4}\int d^6x\sqrt{g} \, h^*_{ab}\left(h^{*ab}\pm h^{ab}
       \right)\pm\int {1\over 2}dBC.
\eea 
$S_0^+$ will describe anti self--dual bosons ($H^+=0$)
and $S_0^-$ self--dual bosons ($H^-=0$). 
The last term, $\int dBC$, is of the Wess--Zumino type and is absent 
for free chiral bosons. 

What selects this form of the action are  essentially 
the local symmetries it possesses. Under a general variation of the fields
$B$ and $a$ it varies, in fact, as
\beq
\label{dS0}
\delta S_0^\pm = \pm2\int \left(vh^\pm d\delta B + 
{v\over \sqrt{-u^2}} h^{\pm}h^{\pm}   d\delta a\right).
\eeq
From this formula it is rather easy to see that $\delta S^\pm_0$ vanishes 
for the following three bosonic transformations,  with transformation
parameters $\Lambda$ and $\psi$, which are one--forms, and $\varphi$ 
which is a scalar:
\bea\label{bos}
&I)&\qquad \delta B=d\Lambda,\qquad \delta a =0\nonumber\\
&II)&\qquad \delta B= -{2h^\pm\over \sqrt{-u^2}}\,\varphi,\qquad
\delta a =\varphi\nonumber\\
&III)&\qquad \delta B=\psi da ,\qquad \delta a =0.
\eea

The transformation $I)$ represents just the ordinary gauge invariance for
abelian two--form gauge potentials.
The symmetry $II)$ implies that $a(x)$ is an 
auxiliary field which does, therefore,  not correspond to a propagating 
degree o freedom\footnote{Notice however that, since the action becomes 
singular in the limit of a vanishing or constant $a(x)$, the gauge 
$d a(x) = 0$ is not allowed.}. Finally, the symmetry $III)$ eliminates
half of the propagating degrees of freedom carried by $B$ and allows to
reduce the second order equation of motion for this field to the desired
first order equation, i.e. \eref{eqm}. To see this we note that the equations
of motion for $B$ and $a$, which  can be read from \eref{dS0}, 
are given respectively by
\bea
d\left(vh^\pm\right)&=&0\label{emb}\\
d\left({v\over \sqrt{-u^2}}h^\pm h^\pm\right)&=&0.
\eea 
First of all it is straightforward to check that the $a$--equation is 
implied by the $B$-equation, as expected, while the general solution of 
the $B$--equation is given by
\beq
\label{sol}
vh^\pm={1\over 2}d\tilde{\psi}da,
\eeq
for some one--form $\tilde{\psi}$. On the other hand under the 
transformation $III)$ we have 
$$
\delta\left(vh^\pm\right)={1\over 2}d\psi da
$$
which is precisely of the same form as \eref{sol}.
Therefore we can use this symmetry to reduce the $B$-equation \eref{emb}
to $vh^\pm=0$ which amounts to $h^\pm=0$, and this equation, in turn, can 
be easily seen to be equivalent to $H^\pm=0$, which is the desired 
chirality condition.

This concludes the proof that the actions $S_0^\pm$ describe indeed
correctly the propagation of chiral bosons.

In a theory in which the $B$ field is coupled to other dynamical 
fields, for example in supergravity theories, we can now conclude
that the complete action has to be of the form 
$$
S=S_0^\pm+S_6,
$$
where $S_6$ contains the kinetic and interaction terms for the fields
to which $B$ is coupled. To maintain the symmetries $I)$--$III)$
one has to assume that those fields are invariant under these
transformations and, moreover, that $S_6$ is independent off the $B$
and $a$ fields themselves.

For more general chirality conditions describing self--interacting 
chiral bosons, like e.g. those of the Born--Infeld type, see ref. \cite{PST3}. 

To conclude this section we introduce two three--form fields,
$K^\pm$, which will play a central role in the next sections
due to their remarkable properties. They are defined as
\beq\label{k}
K^\pm\equiv H+2vh^\mp
\eeq
and are uniquely determined by the following peculiar properties:
\begin{itemize}

\item[i)] they are (anti) self--dual: $K^{\pm*} = \pm K^{\pm}$;
\item[ii)]they reduce to $H^\pm$ respectively  if  $H^\mp= 0$;
\item[iii)] they are invariant under the symmetries $I)$ and $III)$, 
         and under $II)$ modulo the field equations \eref{emb}.

\end{itemize}
These fields constitute therefore a kind of off--shell generalizations
of $H^\pm$.

\section{The action for a free $N=1$, $D=6$ tensor multiplet}

In this section we illustrate the compatibility of the general 
method for chiral bosons with supersymmetry in a 
simple example i.e. the one involving only one free
tensor supermultiplet in flat space--time in six dimensions. The strategy
developed in this case admits natural extensions to more general
cases as we will see in the next two sections.

An $N=1,D=6$  tensor supermultiplet is made out of an antisymmetric tensor 
$B_{[ab]}$, a symplectic Majorana--Weyl 
spinor $\l_{\a i}$ ($\a = 1,\ldots,4; i = 1,2$) and a real scalar $\f$.
For more details on our spinor conventions, see the appendix.
The equations of motion for this multiplet and its on--shell susy transformation
rules are well known. The scalar obeys the free Klein--Gordon equation,
the spinor the free Dirac equation and the $B$--field the self--duality
condition
$$
H^-=0,
$$
where $H=dB$, which means that in this case we have $C=0$.

The on-shell supersymmetry transformations, with rigid transformation
parameter $\xi^{\a i}$, are given by 
\bea
\label{susy}
\delta_\xi \f &=& \xi^i \l_i, \nonumber\\
\delta_\xi \l_{ i} &=& \left( \G^a \D_a \f +
\frac{1}{12} \G^{abc}H_{abc}^+ \right)\xi_i,\nonumber\\
\delta_\xi B_{ab} &=& - \xi^i \G_{ab} \l_i.
\eea
The $USp(1)$  indices 
$i,j$ are raised and lowered according to
$ K_i = \e_{ij} K^j,  K^i = - \e^{ij} K_j, $ where $\e_{12} = 
\e^{12} = -\e_{21} = -\e^{21} = 1 $, \cite{DL}.

Since the equations of motion are free our ansatz for the action, which 
depends now also on the auxiliary field $a$, is 
\beq\label{SH}
S=S_0^-+S_6
=- \int v h^- H +{1\over 2}\int d^6x \left(\l^i \G^a \partial_a \l_i +
\D_a \f \D^a \f \right).
\eeq
This action is invariant under the symmetries $I)$--$III)$ if we assume
that $\f$ and $\l$ are invariant under these transformations.

For what concerns supersymmetry we choose first of all the transformation
for $a$
$$
\delta_\xi a=0,
$$
which is motivated by the fact that $a$ is non propagating and
does therefore not need a supersymmetric partner. Next we should 
find the off--shell generalizations of \eref{susy}. For dimensional
reasons only $\delta_\xi\l$ allows for such an extension. To find
it we compute the susy variation of $S_0^-$, which depends only on $B$ and
$a$, as
$$
\delta_\xi S_0^-=-2\int vh^-d\delta_\xi B=-\int K^+d\delta_\xi B
$$
in which the self-dual field $K^+$, defined in the previous section,
appears automatically. Since $\delta_\xi S_0^-$  should be cancelled by
$\delta S_6$ this suggests to define the off--shell
susy transformation of $\l$ by making the simple replacement
$H^+\ra K^+$, i.e.
$$
\delta_\xi \l_{ i}\ra \bar\delta_\xi \l_{ i} = \left( \G^a \D_a \f +
\frac{1}{12} \G^{abc}K_{abc}^+ \right)\xi_i.
$$
With this modification it is now a simple exercise to show that
the action \eref{SH} is indeed invariant under supersymmetry.
The relative coefficients of the terms in the action are actually
fixed by supersymmetry.

The general rules for writing covariant actions for supersymmetric
theories with chiral bosons, 
which emerge from this simple example, are the following.
First one has to determine the on--shell susy transformations of the 
fields and their equations of motion, in particular one has to determine
the form of the three-form $C$. The off--shell extensions of the susy
transformation laws are obtained by substituting in the transformations
of the fermions $H^\pm\ra K^\pm$. The action has then to be written
as $S_0^\pm+S_6$ where the relative coefficients of the various terms 
in $S_6$ have to be determined by susy invariance. The field $a$, finally,
is required  to be supersymmetry invariant. 
    
An essential step in this procedure is the determination of the susy
transformation laws and equations of motion for the fields. This can 
generally be done most conveniently using superspace techniques, 
especially in the case of supergravity theories, like the ones in the
subsequent sections. Here we illustrate the procedure by
rephrasing the results in \eref{susy} in a (flat) superspace language.
We follow here the superspace conventions of ref. \cite{DL} to which 
we refer the reader for more details on our notations.

One introduces the supervielbeins $e^A=(e^a, e^{\a i}=d\vartheta^{\a i})$, 
which are one--superforms, and the 
three--superform $A = \hat dB$ where now $B$ is a two--{\it super}form 
and $\hat d=e^a\D_a+e^{\a i}D_{\a i} $ 
is the superspace differential. The superspace torsion
is the one--form $T^A=\hat de^A$ and a generic $p$-superform allows a
decomposition analogous to \eref{dec} with $e^a\ra e^A$. 

Then one imposes the rigid superspace constraints
\bea
T^a &=& - e^i   \G^a e_i, \label{vincTa}\\
T^{\a i} &=& 0, \label{vincTai}\\
A_{\a i\b j\g k} &=& 0, \label{vincH} \\
A_{a \a i \b j} &=& - 2 \f \e_{ij} (\G_a)_{\a\b} 
\eea
and solves the Bianchi identity 
\beq
\label{IBH}
\hat d A = 0.
\eeq
The solution gives 
\beq\label{soluz}
\hat dB =A= \frac{1}{3!} e^a   e^b   e^c H_{cba} 
- \left(e^i   \G_a  e_i\right) e^a \f - \frac{1}{2} e^b   e^a
  (e^i \G_{ab} \l_i) 
\eeq
and
\bea\label{spi}
\l_{\a i} &=& D_{\a i} \f,\nonumber \\
D_{\a i} \l_{\b j} &=& \e_{ij} \left( \G^a_{\a\b} \D_a \f - \frac{1}{12} 
(\G^{abc})_{\a\b} H^+_{abc} \right),\nonumber \\
D_{\a i} H_{abc} &=& -3 (\G_{[ab})_\a{}^\b \D_{c]} \l_{\b i},\nonumber\\
H^-_{abc} &=& 0.
\eea

The lowest components (obtained at $\vartheta=d\vartheta=0$)
of the superfields $\f$, $\l_{\a i}$ and $B$ are 
respectively the scalar, spinor and tensor component fields of the
supermultiplet. Notice also that the Bianchi identity
\eref{IBH}  forces the tensor $B$ to be on--shell. In fact, taking
the $\vartheta=d\vartheta=0$ component of \eref{soluz} the last two 
terms go to zero and the last equation in \eref{spi} becomes simply 
$H^-=0$, where now $H\equiv dB$ in ordinary space; 
here we retrieve the result $C=0$.

In superspace formalism the supersymmetry transformation of a superfield
is given by its Lie derivative along the vector field
$\xi = \xi^{\a i} D_{\a i}$. The transformation of the component
fields is just obtained by performing the superspace Lie derivative
and setting finally $\vartheta=d\vartheta=0$.
Using this one can read the transformations \eref{susy} directly from
the spinorial derivatives given in \eref{spi}.

\section{The action for pure $N = 1$, $D=6$ supergravity}

The supergravity multiplet in six dimensions contains the graviton, 
a gravitino and an 
antisymmetric tensor with anti--selfdual (generalized) field strength. 
The graviton is described by the vector--like vielbein $e^a = dx^m 
{e_m}^a$, the gravitino by the spinor--like one--form $e^{\a i} = 
dx^m {e_m}^{\a i}$ and the tensor by the two--form $B$. We  
introduce also the Lorentz connection one--form 
$\omega^{ab} = dx^m \omega_{m}{}^{ab}$ 
in order to implement a first order formalism.
As outlined at the end of 
the previous section we promote these forms to superforms and 
define super--torsions and super--curvatures as
\bea
T^a &\equiv& D e^a = \hat d e^a + e^{b}   {\omega_b}^a,\nonumber\\
T^{\a i} &\equiv& D e^{\a i} = \hat d e^{\a i} + \frac{1}{4} e^{\b i} 
  (\G_{ab})_\b{}^\a \omega^{ab}, \nonumber\\
R^{ab} &\equiv& \hat d \omega^{ab} +  \omega^{ac}   {\omega_c}^{b}\nonumber\\
A&\equiv& \hat d B.
\eea
The standard constraints read now
\bea
T_{\a i\b j}{}^a &=&-2\e_{ij}\G^a_{\a\b} \\
T_{ab}{}^c &=& 0\\
A_{\a i\b j\g k} &=& 0 \\
A_{a \a i \b j} &=& - 2 \e_{ij} (\G_a)_{\a\b}, 
\eea
and the solution of the relevant superspace Bianchi identities
leads to the following parametrizations of these torsions and curvatures:
\bea
\label{vinc1}
T^a &=& - e^i\G^a e_i,\\
T^{\a i} &=&  \frac{1}{8} e^{\b i} e^a (\G^{bc})_\b{}^\a 
H^-_{abc} + \frac{1}{2} e^{a} e^b T_{ba}{}^{\a i}, \label{vinc2}\\
R_{ab} &=& \frac{1}{2} e^i \G^c e_i  
H^-_{abc} - e^{\a i}   e^c [ (\G_c)_{\a\b} T_{ab}{}^\b_i - 2 
(\G_{[a})_{\a\b} T_{b]c}{}^\b_i]  \label{vinc3}
+ \frac{1}{2} e^d e^c R_{cdab}, \\
\hat dB =A&=& \frac{1}{3!} e^a e^b e^c H_{cba} 
- \left(e^i\G_a  e_i\right) e^a.\label{vinc4} 
\eea

The constraint of a vanishing purely bosonic torsion $T_{abc}$ 
is, in general,  conventional in that, through a redefinition of the 
connection, it can be set equal to any tensor. Sometimes it
is more convenient, in fact, to have $T_{abc}\neq 0$, see e.g.
\cite{DL}. In the present case, however, having $T_{abc}=0$
constrains $\omega_{m}{}^{ab}$ to be the usual super--covariant
connection which depends only on the graviton and the gravitino;
it has, in particular, no spurious dependence on $B$. This implies,
in turn, that the covariant derivatives are automatically invariant
under the bosonic transformations $I)$--$III)$ and, 
to maintain those symmetries, one has to avoid in
$S_6$ only the explicit appearance of $B$. 

The $A$--Bianchi identity implies, in particular, also the 
$B$--equation of motion
$$
H_{abc}^+=0.
$$
Due to \eref{vinc4} this implies, in ordinary space--time, $H^+=0$,
where now
$$
H=dB+ \left(e^i\G_a  e_i\right) e^a= \frac{1}{3!} e^a e^b e^c H_{cba}. 
$$
This means that in this case the three--form $C$ is non vanishing
being given by
\beq\label{C3}
C=\left(e^i\G_a  e_i\right) e^a.
\eeq

The supersymmetry transformations of $e^a$, $e^{\a i}$, $\omega^{ab}$ and 
$B$ can 
obtained as (covariant) Lie derivatives of these forms along the local 
superspace vector $\xi(x) = \xi^{\a i} (x) D_{\a i}$ and, therefore,
they can be read directly from \eref{vinc1}--\eref{vinc4}:
\bea
\label{susyi}
\delta_\xi e^a &=& i_{\xi} T^a = -2 \xi^i \G^a e_i, \\
\delta_\xi e^{\a i} &=& D \xi^{\a i} + i_\xi T^{\a i} = D 
\xi^{\a i} - \frac{1}{8} \xi^{\b i}  e^a (\G^{bc})_\b{}^\a 
K^-_{abc}, \label{28b}\\
\delta_\xi \omega_{ab} &=& i_\xi R_{ab} = \xi^i \G^c e_i K^-_{cab} - 
\xi^{\a i} e^c \left( (\G_c)_{\a\b} T_{ab}{}^\b_i - 2 (\G_{[a})_{\a\b} 
T_{b]c}{}^\b_i\right), \\
\delta_\xi B &=& i_\xi A = -2 (\xi^i \G_a e_i) e^a,\\
\delta_\xi a &=& 0.
\eea

In these relations, according to our general rule, we made already the 
replacements $H^-_{abc}\ra K^-_{abc}$ -- which occur, in fact, only
in the gravitino transformation in that the connection $\omega_{ab}$ is not
an independent field but depends on the graviton and the gravitino -- 
and we added the trivial 
transformation law for the auxiliary field $a$. 

As it stands, this trivial
transformation law does not seem to preserve the susy algebra in that
the commutator of two supersymmetries does not amount to a translation.
On the other hand it is known that the supersymmetry algebra closes on 
the other  symmetries of a theory; in the present case it is easily seen 
that the anticommutator of two susy transformations on the $a$ 
field closes on the  bosonic transformations $II)$. It is also 
interesting to observe that the new terms in the susy transformations 
of the gravitino  are exactly the ones that ensure the closure of the total
symmetry algebra on the $B$--field, and again susy closes on the 
transformations $II)$\footnote{We thank D.Sorokin for these remarks.}.

The covariant action for pure $N = 1$, $D = 6$ supergravity can 
now be written as 
\bea\label{azsu} 
S&=&S_0^+ +\int L_6\\
     S_0^+&=& \int \left(v h^+ H + {1\over 2} dB C\right)\\
     L_6 &=&
 \frac{1}{48} \e_{a_1 \ldots a_6 } e^{a_1} e^{a_2} e^{a_3} e^{a_4} 
R^{a_5a_6}  - \frac{1}{3} e^{a_1}e^{a_2}e^{a_3} (De^i\G_{a_1 a_2 a_3}e_i). 
\eea
The three--form $C$ is given in eq. \eref{C3} and for convenience
we wrote the term $S_6$ as an integral of a six--form, $L_6$. This
six--form contains just the Einstein term, relative to the 
super--covariantized spin connection, and the kinetic term for
the gravitino. The relative coefficients are fixed by susy invariance,
see below. In this case $S_0^+$ contains also the couplings of
$B$ to  the gravitino and the graviton.

This action is invariant under the symmetries $I)$--$III)$ because
$L_6$ does not contain $B$ and we assume the graviton and the gravitino
to be invariant under those transformations.

The evaluation of the supersymmetry variation of $S$ is now a merely
technical point. The variation of $\int(vh^+H)$ has to be computed
"by hand" while the variation of the remaining terms, which are forms,
is most conveniently evaluated by lifting them to superspace,
performing their superspace differential and taking the interior
product with the vector $\xi$. The results are
\bea\label{var}
\delta_\xi S_0^+&=&
\int \left(i_\xi R + \frac{1}{2} (\xi^i \G^a e_i)  e^b 
 e^c K^-_{cba}\right)K^-  -{1\over 2} \int i_\xi\left(RC\right)\\
\delta_\xi \int L_6 &=& \int i_\xi \hat dL_6=
\int i_\xi\left({1\over 2}RC-{1\over 3} e^a  e^b  e^c (T^i  \G_{abc} T_i)
\right).
\eea
Here we defined
$$
R=\hat d C=2 (T^i  \G_a e_i) e^a,
$$
and the parametrization of $T^i=De^i$ is given in \eref{vinc2}
(with $H_{abc}^-\ra K_{abc}^-$). 

We see that the susy variation of $S_0^+$ depends on $B$ only through 
the combination $K^-$, justifying again our rule for the modified susy 
transformation rules for the fermions.

The susy variation of the total action becomes then
$$
\delta_\xi S=
\int \left(i_\xi R + \frac{1}{2} (\xi^i \G^a e_i)  e^b 
 e^c K^-_{cba}\right)K^-  -{2\over 3} e^a  e^b  e^c 
(i_\xi T^i) \G_{abc} T_i
$$
and is easily seen to vanish using the expression for $i_\xi T^i$
given in \eref{28b}. 

\section{$N = 1$ supergravity coupled to $n$ tensor multiplets}

As last example we consider the case in which the supergravity 
multiplet is coupled to an arbitrary number $n$ of tensor multiplets. 
This situation arises, for example, in $M$--theory compactified on
$(K_3\times S^1)/Z_2$ \cite{Sen} 
and, until now, for this system a covariant lagrangian
formulation did not exist. The purpose of this section is to fill this 
gap. For simplicity we will disregard vector and hypermultiplets.

The  field equations for supergravity coupled to an arbitrary number
of tensor multiplets have been obtained in \cite{sugra6d}. For a recent 
account see ref. \cite{new6d}, where this system has been generalized to 
include 
hypermultiplets and vector multiplets, see also \cite{sa}.
For what concerns the geometrical 
aspects involved in this system we will
follow basically the notations of \cite{new6d}.

The supergravity multiplet is described as before by the vielbein one--form 
$e^a$, the gravitino one--form $e^{\a i}$ and by the two--form 
$B$. The $n$ tensor supermultiplets are composed by $n$
two--forms $B^{\um}$, $n$ symplectic Majorana--Weyl spinors 
$\l^{\um}_{\a i}$ and by $n$ scalars $\f^{\ua}$ ($\ua = 1,\ldots,n$) 
which parametrize as local coordinates the coset space $SO(1,n)/SO(n)$. 
The indices $\um = (1,\ldots,n)$ span the fundamental representation 
of $SO(n)$. The two--form of the supergravity multiplet and the $n$
two--forms of the tensor multiplets are collectively denoted by
$B^I$, where the indices $I,J = (0,1,\ldots, n)$ span the 
fundamental representation of $SO(1,n)$. 

A convenient way to parametrize
the coset geometry is to introduce the $SO(1,n)$ group element
$L(\f)\equiv (L^I, L^I_{\um})$ and its inverse $L^{-1}(\f) 
\equiv (L_I, L_I^{\um} )$, which are local functions on the coset space
and obey
\beq
\label{LL}
L^I L_I = 1, \quad L^I_{\um} L_I = 0 = L^I L_I^{\um}, \quad L^I_{\um}
 L_I^{\un} = \delta_{\um}^{\un}.
\eeq
The coset connection $A_{\ua}{}^{\um}{}_{\un}(\f)$ and the coset vielbein
$V_{\ua}{}^{\um}(\f)$ are defined by
\beq
 D_{\ua} L_I^{\um}\equiv \D_{\ua} L_I^{\um} + A_{\ua}{}^{\um}{}_{\un}
L_I^{\un}  = {V_{\ua}}^{\um} L_I,
\eeq
\beq
\label{DL}
\D_{\ua} L_I = {V_{\ua}}^{\um}  L_{I\um}.
\eeq
These relations imply also that 
\beq
\label{defeta}
\eta_{IJ} = -L_I L_J + L_I^{\um}L_{J\um}
\eeq
is the constant Minkowski metric of $SO(1,n)$.
It is also convenient to define the matrix
\beq
\label{defG}
G_{IJ}(\f) = L_I L_J + L_I^{\um}L_{J\um}
\eeq
which is a local function on the coset. 

The above relations imply, in 
particular, that the coset manifold is a maximally symmetric space
with constant negative curvature. This property of the manifold ensures, 
ultimately,
the supersymmetry of the equations of motion and the closure of the
susy algebra \cite{new6d}.

It is also convenient to introduce for any $SO(1,n)$ vector $W^I$
its components $W=L_IW^I$ and $W^{\um}=L_I^{\um}W^I$.

The equations of motion and susy transformations can now be derived
through the superspace techniques outlined in the previous sections.
In addition to the super--torsion we introduce $n+1$ two--superforms
$B^I$ and their supercurvatures $A^I=\hat dB^I$ and impose the constraints
\bea
T_{\a i\b j}{}^a &=&-2\e_{ij}\G^a_{\a\b} \\
T_{ab}{}^c &=& 0\\
A_{\a i\b j\g k}^I &=& 0 \\
A_{a \a i \b j}^I &=& - 2 \e_{ij}L^I (\G_a)_{\a\b}.
\eea
The solution of the torsion Bianchi identities and of $\hat dA^I=0$
leads now to the parametrizations
\bea
\label{56}
T^a &=& - e^i  \G^a e_i,\\
T^{\a i} &=&    e^{\b j} e^a \left( 3 \delta_\b{}^\a 
V_{a\,j}{}^i + (\G_{ab})_\b{}^\a V^b{}_{j}{}^i + \frac{1}{8} 
\delta_j{}^i (\G^{bc})_\b{}^\a A^-_{abc} \right) + \frac{1}{2} e^a  
e^b T_{ba}{}^{a i},\\
R_{ab} &=& \frac{1}{2} e^{\a i}  e^{\b j} \left( \e_{ij} 
(\G^c)_{\a\b} A^-_{cab}  - 4 (\G_{abc})_{\a\b} V^c_{ij}\right)  \nonumber\\
&-&  e^{\a i}e^c \left((\G_c)_{\a\b} T_{ab}{}^\b_i - 2 (\G_{[a})_{\a\b} 
T_{b]c}{}^\b_i\right)
 + \frac{1}{2} e^d  e^c R_{cdab},\\
\hat dB^I=A^I&=&\frac{1}{3!} e^a e^b e^c A_{cba}^I 
- \left(e^i\G_a  e_i\right) e^a L^I
+{1\over 2}e^be^a e^i(\G_{ab})\l_i^{\um}L^I_{\um}  \label{defh},\\
D \f^{\ua} &=& e^i \l_i^{\um} V_{\um}{}^{\ua}  + e^a D_a 
\f^{\ua}, \\
D\l_{\a i}^{\um} &=& -e^\b_i\left((\G^a)_{\b\a} D_a \f^{\ua} 
V_{\ua}{}^{\um} + \frac{1}{12} (\G^{abc})_{\a\b} 
A^{\um +}_{abc}\right)  + e^a D_a \l_{\a i}^{\um}. \label{62}
\eea
Here the $n$ fermions in the tensor multiplet are represented by
the superfields $\l_{\a i}^{\um}=D_{\a i}\f^{\ua}V^{\um}_{\ua}$
and we defined  
\beq
V^a_{ij} = - \frac{1}{16} \l_{i}^{\um} \G^a \l_{j\um}.
\eeq
According to our convention above we have also $A_{abc}=L_IA_{abc}^I$ 
and $A^{\um}_{abc}=L_I^{\um} A^I_{abc}$.

Most importantly, the closure of the susy algebra imposes also
the (anti)-selfduality equations of motion
\bea\label{asd}
A^{\um-}_{abc}&=&0\\
A^+_{abc}&=&{1\over8}\l^i_{\um}(\G_{abc})\l_i^{\um}.
\eea
Due to \eref{defh} these equations amount to 
\bea\label{s1}
H^{\um -}&=&0\\
H^+&=&0,\label{s2}
\eea
where the generalized curvatures $H^I$ are defined, now in ordinary space, 
as
\bea
H^I&=&dB^I+C^I,\\
C^I&=&\left((e^i  \G_a e_i)  e^a - \frac{1}{48} e^a  e^b  
e^c ( \l^i_{\um} \G_{cba} \l_i^{\um})\right)L^I
-{1\over 2}e^a  e^b  (e^i \G_{ba} \l_i^{\um})L^I_{\um}\label{ci}\\
&\equiv& CL^I+C^{\um} L^I_{\um}. \nonumber
\eea

Introducing now again a single auxiliary field $a$ and the
related vector $v^a$ as in section two, we define the
$n+1$ two--forms $h^I$, which replace the two--forms $h^\pm$ 
introduced in the preceding sections, as
\bea\label{hpicc}
h^I&=&{1\over 2}e^ae^b h^I_{ba}\\
h^I_{ba}&=&{1\over 2}v^c\left(H_{cba}^I-\eta^{IK}G_{KJ}H_{cba}^{J*}
        \right)\label{hpicc1}\\
         &=&v^c\left(L^IH_{cba}^++L_{\um}^IH^{\um-}_{cba}\right).
        \label{hpicc2}  
\eea
The three--forms $K^I$, corresponding to $K^\pm$, are then given by 
$$
K^I=H^I+2vh^I.
$$
Their components along $L_I$ and $L_I^{\um}$
satisfy identically the (anti)--selfduality conditions
$$
K^{*}=-K, \qquad K^{\um*}=K^{\um}
$$
and reduce respectively to $H^{-}$ and  $H^{\um+}$ if 
the (anti)--selfduality conditions \eref{s1} and \eref{s2}
are satisfied.
The off--shell susy transformations are now obtained from
\eref{56}--\eref{62} in the usual way with the replacements
\beq
\label{RP}
A_{abc}^-\ra K_{abc},   \qquad A^{\um+} _{abc}\ra K^{\um}_{abc}.
\eeq
The covariant action for this system can now be written
again in the form 
\beq
S = S_0 +\int L_6
\eeq
where
\bea\label{Sn}
S_0&=&-\int\eta_{IJ}\left(vh^IH^J+{1\over 2}dB^I C^J\right)\\
   &=& \int \left(v h  H+ {1\over 2}  HC\right)
    -\left(vh_{\um} H^{\um} +{1\over 2}H_{\um} C^{\um}\right)
\eea
and
\bea
L_6 &=& 
\frac{1}{48} \e_{a_1 \cdots a_6 } e^{a_1} e^{a_2} e^{a_3} e^{a_4} 
R^{a_5a_6}  - \frac{1}{3} e^{a_1}e^{a_2}e^{a_3} (De^i\G_{a_1 a_2 a_3}e_i)
+ \nonumber\\
&-& \frac{1}{2} \frac{\e_{a_1 \cdots a_6}}{5!} 
e^{a_1} \cdots  e^{a_5} (\l^{i}_{\um} \G^{a_6} D \l_{i}^{\um}) 
+ \nonumber\\
&+& \e_{a_1 \cdots a_6} \left( \frac{e^{a_1} \cdots  e^{a_6} }{6!}
 \frac{1}{2} Q_{b\ua} Q^{b\ua} - 
 \frac{e^{a_1} \cdots  e^{a_5} }{5!} (D \f^{\ua}- 
e^i \l_i^{\um} V_{\um}{}^{\ua}) Q_{\ua}^{a_6} \right) + \nonumber \\
&-& \frac{1}{2} 
\frac{\e_{a_1 \cdots a_6}}{4!} e^{a_1} \cdots  e^{a_4}  
(e^i \G^{a_5 a_6} \l_{i\um})  (D \f^{\ua} 
V_{\ua}{}^{\um} - e^j \l_j^{\um})+ \nonumber \\
&-& 16 \frac{\e_{a_1 \cdots 
a_6}}{6!} e^{a_1}  \cdots  e^{a_6} V_{b\,ij} V^{b ij} - 
\frac{1}{3} e^{a_1}  \cdots  e^{a_4}  (e^i \G_{a_1 a_2 
a_3}  e^j) V_{a_4 ij}.
\eea

In $L_6$ there are no quartic terms in the 
gravitino because the unique term which respects 
all the symmetries would be $T^a T^b E_a E_b$ and this vanishes
due to the cyclic gamma matrix identity in six dimensions. 
The relative coefficients of the various terms are fixed by
susy, see below.
The field $Q_{b\ua}$ is an auxiliary  field which has been introduced
in order to write $L_6$ as a six--form. Its equation of motion gives
$e^bQ_{b}^{\ua}=D\f^{\ua}- e^i\l_i^{\um}V^{\ua}_{\um}$ and, upon substituting
this back in $\int L_6$, one obtains the usual super--covariantized
kinetic term for the scalars in the tensor multiplet.

Since under a generic variation of $B^I$ and $a$ one has
\beq\label{dds}
\delta S_0 = -2\int\eta_{IJ} \left(vh^I d\delta B^J + 
{v\over \sqrt{-u^2}} h^I h^J d\delta a\right),
\eeq
$S_0$ is now evidently invariant under the bosonic symmetries
$I)$--$III)$ which in the present case take the form
\bea
&I)&\qquad \delta B^I=d\Lambda^I,\qquad \delta a =0\\
&II)&\qquad \delta B^I= -{2h^I \over \sqrt{-u^2}}\,\varphi,\qquad
\delta a =\varphi\\
&III)&\qquad \delta B^I=\psi^I da ,\qquad \delta a =0.
\eea
Under these transformations  $\int L_6$ is trivially invariant.

From \eref{dds} one sees that the equations of motion for $B^I$ become 
in this case  $d(vh^I)=0$ which, by fixing the symmetries $III)$,
can be reduced to $h^I=0$ and these last equations correspond just to 
\eref{s1} and \eref{s2} (see \eref{hpicc2}). The $a$--equation of motion
is again a consequence of the $B^I$--equations.

The supersymmetry variation of the action can be computed as in
the previous section:
\bea
\delta_\xi S_0&=&
-\int \eta_{IJ}  \left(i_\xi R^I + \frac{1}{2} (\xi^i \G^a e_i)  e^b 
 e^c K^I_{cba}\right)K^J+\nonumber\\
&+&{1\over 2} \int i_\xi \left(\eta_{IJ}R^IC^J\right)+
(\xi^i \l_i^{\um}) K  K_{\um},\label{var0}\\
\delta_\xi \int L_6 &=& \int i_\xi \hat dL_6.
\eea
Here we defined the four--forms
$$
R^I=\hat d C^I.
$$

The variation of $S_0$ depends on $B^I$ again only through
$K^I$ and, with respect to the expression found in \eref{var}, there
is an additional term, proportional to $\lambda_{\um}$, which comes 
from the self--interactions of the tensor multiplet. The action $S_0$
depends on the geometry of the coset manifold, in fact, only 
through the matrix $G_{IJ}$ (see eqs. \eref{hpicc},\eref{hpicc1}). Since
\beq
\hat d G_{IJ} =2D \f^{\ua} V_{\ua\um} (L_I^{\um} L_J + L_I L_J^{\um})
\eeq
we have 
\beq
\delta_\xi G_{IJ} = 2 \xi^i \l_i^{\um} (L_{I\um} L_J + L_I L_{J\um})
\eeq
and this leads to the additional term in $\delta_\xi S_0$.

The explicit expressions for  $R^I=\hat d C^I$ and $\hat dL_6$
can be obtained using \eref{56}--\eref{62}, with the replacements
\eref{RP}, and 
with a long but straightforward calculation one can show that
$\delta_\xi S$ indeed vanishes. The explicit expression for
$R^I$ can be found in the appendix. 

\section{Concluding remarks}

In this paper we analyzed, in the framework of the approach of 
\cite{PST,PST3,M5}, supersymmetric and supergravity models 
with chiral bosons in six dimensions, and, outlining a general
procedure, we wrote covariant and supersymmetric actions for some 
of these models.   

All these actions have the structure of \eref{S0} and are invariant 
under the bosonic symmetries $I)$--$III)$, typical of the approach of 
\cite{PST}-\cite{PSTprd}, as 
well as under (modified) supersymmetry transformations. Our general recipe is
that these modified transformations are obtained from the standard 
ones in a natural and universal way by simply replacing the 
(anti)--selfdual field strength tensors $H^{(\pm)}_{abc}$, which arise in the 
standard supersymmetry transformations of the fermions, 
with the special tensors $K^{(\pm)}_{abc}$. 

We treated in detail the case of a free tensor supermultiplet, pure 
$N = 1$ supergravity and $N = 1$ supergravity coupled with $n$ 
tensor supermultiplets. More general models containing  also 
hypermultiplets and vector multiplets have not been considered 
explicitly in this paper. The inclusion of an arbitrary number 
of hypermultiplets does not present any conceptual new
difficulty; we did not consider them here only for the sake 
of simplicity.

The same holds for the inclusion of Yang--Mills supermultiplets \cite{sa}
a part from 
the following important observation. In this case, in fact, the
tensor multiplets can be coupled to a certain number of Lie--algebra
valued Yang--Mills
fields $A_k$, with curvatures $F_k=dA_k+A_kA_k$, where $k=1,\cdots,N_{YM}$
and $N_{YM}$ is the number of factor gauge groups. The couplings are
realized in a standard way by defining the generalized curvatures now
as
\beq\label{hym}
H^I=dB^I+\tilde C^I + c^I_k \, \omega_{YM}(A^k),
\eeq
where $d\omega_{YM}(A_k)=tr(F^kF^k)$ defines the usual Chern--Simons
three--forms, one for each factor group, and $\tilde C^I$ is given by 
$C^I$ in \eref{ci} augmented by a term proportional to the gluino
bi--linears, see e.g. \cite{new6d}. The $c^I_k$ form a {\it constant} 
matrix, which weighs the couplings of the various Chern--Simons terms
to the tensor multiplets,  
and in \eref{hym} a sum over $k$ is understood.
The (self)--duality  equations of motion for these $H^I$ are 
then again given by \eref{s1},\eref{s2}. 

The appearance of the 
Chern--Simons terms leads then in the action $S_0$ in \eref{Sn}
to a contribution given by
\beq\label{GS}
-{1\over 2}\int \eta_{IJ}\, dB^I c^J_k \omega_{YM}(A^k) 
\eeq
which, in general, is not gauge invariant its variation leading to 
\beq
\delta_{YM}S_0=
{1\over 2}\int \eta_{IJ}c^I_k c^J_ltr(\lambda^ldA^l)tr(F^kF^k),
\eeq
where the $\lambda^l$ are the Yang--Mills transformation parameters.
For cohomological reasons this would then imply that the action
necessarily breaks also supersymmetry. 

To save gauge invariance, and
also supersymmetry, one has to impose that the constants which weigh
the various Chern--Simons terms in \eref{hym} have to be constrained
by
\beq\label{const}
\eta_{IJ}c^I_k c^J_l=0.
\eeq
If this relation is not satisfied one has a set of supersymmetric and
gauge invariant equations of motion which are non integrable, in the sense 
that they can not be deduced from 
an action. In such a situation  conservation of energy--momentum
and of the Yang--Mills currents is not guaranteed. In ref. 
\cite{new6d} it was, indeed, found that the  Yang--Mills
currents are conserved if and only if the constraint \eref{const} holds.
It can also be seen that, if this constraint holds, then 
the Yang--Mills equation derived from our covariant Lagrangian
coincides with the one given in \cite{new6d} upon fixing the
symmetry $III)$ according to $h^I=0$.

This situation is not new. It was, in particular, noticed in
\cite{DL} that in the case of tensor multiplets coupled
in flat six--dimensional space--time to Yang--Mills supermultiplets
through Chern--Simons terms
one obtains a set of supersymmetric and gauge invariant equations
of motion which do, however, not admit an action. This system represents,
in some sense, a limiting case in which the supergravity multiplet
decouples. In this "limit", however, the geometry of the $n$ tensor
multiplets becomes trivial and the constraint \eref{const} would
reduce to $\delta_{IJ}c^I_k c^J_l=0$, where now $I,J=(1,\cdots,n)$,
whose unique solution is $c_k^I=0$. This means that the flat 
tensor--Yang--Mills system is consistent only in the absence of 
Chern--Simons couplings, i.e. when both systems are free,  
in which case the action becomes simply a sum of $n$ terms like 
the ones given in \eref{SH} plus the free super--Yang--Mills
action.

In the case when the supergravity multiplet is present the meaning 
of the constraint \eref{const} can be understood as follows. Its
general solution is in fact
$$
c_k^I=\alpha_k c^I,
$$
where the $\alpha_k$ are $N_{YM}$ arbitrary but non vanishing constants
and $c^I$ is an $SO(1,n)$ null vector,
$$
\eta_{IJ}c^I\,c^J=0.
$$
This means that there is, actually, a unique total Chern--Simons 
three--form appearing in the theory, given by
$$
\omega_{YM}=\sum_{k=1}^{N_{YM}}\a_k\, \omega_{YM}(A^k),
$$
and the generalized curvatures become then
$$
H^I=dB^I+\tilde C^I + c^I \omega_{YM}.
$$
Choosing for $c^I$ a standard representative, for example
$c^I=(1,0,\cdots,0,1)$, it is easily seen that there is only {\it one}
two--form which carries a Chern--Simons correction, i.e. $B^0+B^n$, 
while all the others, i.e. $B^0-B^n$ and $B^I$ with $1\leq I \leq n-1$,
are gauge invariant. In the case of one tensor multiplet $(n=1)$ 
this means that you can form a two--form with closed field strength
and one whose curvature carries a Chern--Simons correction, the
two field strengths being related on--shell by Hodge duality. It is
indeed known, see e.g. \cite{new6d},\cite{DL}, that for $n=1$ only
under these circumstances you can construct a local, gauge invariant,
and supersymmetric theory. 

If \eref{const} is not satisfied one can interpret 
$\delta_{YM}S_0$ as a "classical" gauge anomaly which, as observed
above, would then also have a supersymmetric partner
\cite{anom}. For an appropriate content of fields it can then happen 
that these  "classical"  anomalies are cancelled 
by the one loop quantum ABBJ anomalies through a generalized
Green--Schwarz mechanism which invloves all the $B^I$ fields \cite{sa},
since \eref{GS} has precisely the structure of the 
Green--Schwarz counterterm. In this case, however, since supersymmetry
and gauge invariance require also the inclusion of one--loop
quantum corrections, it is no longer meaningful to search
for a {\it local} invariant action 
(see also the discussion in\cite{new6d}).

\paragraph{Acknowledgements.}
\ We are grateful to I. Bandos, P. Pasti and D. Sorokin for their interest 
in this work and useful discussions. This work was supported by the 
European Commission TMR programme ERBFMPX-CT96-0045 to which K.L. and 
M.T. are associated.

\section{Appendix}
 
In this appendix we give some details on our notations and conventions 
and report the explicit expression for the four forms $R^I$.

We write the six--dimensional  symplectic Majorana-Weyl spinors as 
$\psi_{\a i}$ 
(left-handed) and $\psi^{\a i}$ (right-handed) where $i = 1,2$ is 
an $USp(1)$ index which can be 
raised and lowered with the invariant antisymmetric tensor $\e_{ij}$
\beq
\psi_i = \e_{ij} \psi^j, \qquad \psi^i = \e^{ji} \psi_j,
\eeq
while $\a=1,\cdots,4$ is a chiral $SO(1,5)$ spinor index which cannot be raised 
or lowered. The symplectic Majorana-Weyl condition reads
\beq
\e^{ij} \psi^{\a j} = O^{\a\b} \psi^{\star \b i} 
\label{A1}
\eeq
where the matrix $O$ satisfies
\beq
O^T = -O, \quad O^\star = O, \quad O^2 = -1.
\eeq
The $4\times 4$ matrices $(\G^{a})_{\a\b}$ and $(\G^a)^{\a\b}$ span a 
Weyl-algebra,
$(\G_{(a})_{\a\b} (\G_{b)})^{\b\g} = \eta_{ab} \delta_\a^\g$,
and satisfy the hermiticity condition
\beq
O \G^a{}^\dagger O = \G^a.  \label{A2}
\eeq
Since our formalism is manifestly $USp(1)$ invariant the relations 
\eref{A1}--\eref{A2} need, however, never be used explicitly.

The duality relations for the anti-symmetrized $\G$-matrices is
\beq
(\G_{a_1\ldots a_k})_{\a\b} = - (-1)^{k(k+1)/2} \frac{1}{(6-k)!} 
\e_{a_1\ldots a_6} \left(\G^{a_{k+1}\ldots a_6}\right)_{\a\b}
\eeq
where no "$\g_7$" appears since our $\G$-matrices are $4\times 4$ 
Weyl matrices. The cyclic identity reads
\beq
(\G^a)_{\a(\b} (\G_a)_{\g)\delta} = 0,
\eeq
and another fundamental identity is
\beq
(\G_a)_{\a\b}(\G^a)^{\g\delta} = -4 \delta^{\g}_{[\a} \delta^\delta_{\b]}.
\eeq

The explicit expression for the four--forms $R^I=\hat d C^I$, 
appearing in the variation \eref{var0}, is 
\bea
R^I&=&-{1\over 2}e^{a_1}e^{a_2} (e^i\G^{a_3}e_i)K^I_{a_1a_2a_3}\\
   &+&\left[e^{a_1}e^{a_2}e^{a_3}e_i\G_{a_1}T^i_{a_2a_3}
    +{3\over 8}e^{a_1}e^{a_2}e^{a_3}e^i\G_{a_1a_2}\l_{i\um}V_{\ua}^{\um}
    D_{a_3}\f^{\ua}\right.\\   
   &-&{1\over 48}\e_{a_1\cdots a_6}e^{a_1}e^{a_2}e^{a_3}e^i\G^{a_4a_5}
\l_{i\um}V_{\ua}^{\um}D^{a_6}\f^{\ua}
-{1\over8}e^{a_1}e^{a_2}e^{a_3}(e^i\G^{a_1}{}_b \l_{i\um})K^{\um}_{ba_2a_3}\\
&+&\left.{1\over 4}e^i\l_{i\um}K^{\um}
+2e^{a_1}e^{a_2}e^{a_3}e^{a_4}D_{a_1}\l_{a_2a_3a_4}\right]L^I+\\
&+&\left[-{1\over 2}e^{a_1}e^{a_2}e^{a_3}e^i\G_{a_1a_2}D_{a_3}\l_i^{\um}
+{1\over 2}e^{a_1}e^{a_2}T^i\G_{a_1a_2}\l_i^{\um}\right.\\
&+&\left.2e^{a_1}e^{a_2}e^{a_3}V_{\ua}^{\um}D\f^{\ua}\l_{a_1a_2a_3}\right]
L_{\um}^I,
\eea
where $\l_{abc}\equiv-{1\over 96}\l_{\um}^i(\G_{abc})\l^{\um}_i$.

\end{document}